\documentclass[prb,groupedaddress]{revtex4}
\usepackage[pdftex]{graphicx,epsfig}

\begin{document}

\title{Equivalent dynamical complexity in a\\ many-body 
quantum and collective human system}

\author{Neil F. Johnson$^1$, Josef Ashkenazi$^1$, Zhenyuan Zhao$^1$ and Luis Quiroga$^2$} 
\affiliation{$^{1}$Physics Department, University of Miami, Coral Gables, FL 33126, U.S.A.\\
$^{2}$Physics Department, Universidad de Los Andes, Bogota, Colombia}

\begin{abstract}
Proponents of Complexity Science believe that the huge variety of emergent phenomena observed throughout nature, are generated by relatively few microscopic mechanisms. Skeptics however point to the lack of concrete examples in which a single mechanistic model manages to capture relevant macroscopic {\em and} microscopic properties for two or more distinct systems operating across radically different length and time scales.  Here we show how a single complexity model built around cluster coalescence and fragmentation, can cross the fundamental divide between many-body quantum physics and social science. It simultaneously (i) explains a mysterious recent finding of Fratini {\it et al.} concerning quantum many-body effects in cuprate superconductors (i.e. scale of $10^{-9}-10^{-4}$ meters and $10^{-12}-10^{-6}$ seconds), (ii) explains the apparent universality of the casualty distributions in distinct human insurgencies and terrorism (i.e. scale of $10^{3}-10^{6}$ meters and $10^{4}-10^{8}$ seconds), (iii) shows consistency with various established empirical facts for financial markets, neurons and human gangs and (iv) makes microscopic sense for each application. Our findings also suggest that a potentially productive shift can be made in  Complexity research toward the identification of equivalent many-body dynamics in both classical and quantum regimes.  
\end{abstract}

\maketitle

The central limit theorem predicts that a collection of objects in which there are no hidden correlations (e.g. a handful of coins being tossed) will  produce fluctuations which have an approximate Gaussian distribution. By contrast, the fluctuations emerging from systems containing correlations which cross multiple length and/or time scales can exhibit significant deviations from Gaussian behavior. A statistical form which emerges from many distinct systems is the exponentially truncated power-law distribution $p(x) \propto x^{-\alpha} {\rm exp}(-x/x_0)$ where $\alpha$ typically takes values between one and four\cite{Gabaix,Stanley,Bohorquez,Barabasi,chialvo,Richardson}. It is tempting to infer that if two such systems $A$ and $B$ share similar values $\alpha_A$ and $\alpha_B$, then they also share a common underlying dynamical mechanism. But this is not generally true. Indeed one of the most damaging criticisms of Complexity Science as a unifying discipline, is that mechanisms which make sense in the context of system $A$ may make little sense for system $B$. For example, a two-dimensional model with nearest-neighbor interactions seems reasonable for vehicle traffic or fish shoals, but not for herding in global financial markets where interactions become essentially independent of spatial separation\cite{Ball,Buchanan,jasny,Cho}. 

Complex phenomena appear across all length and time scales. In the physics community, the toughest complexity arguably lies at the level of quantum many-body phenomena, such as high temperature superconductivity in the cuprates\cite{Fratini}. In the social sciences, it arguably lies in the field of human conflict\cite{Richardson}. A concrete quantitative connection between these two phenomena would seem highly unlikely -- yet it is precisely this connection that we uncover here. Specifically, we use a single dynamical model to simultaneously provide a first explanation for the remarkable yet mysterious recent finding of Fratini {\it et al.}\cite{Fratini} (see Fig. 1(a)) concerning the complex, nanoscale quantum-mechanical world of high-temperature superconductivity, {\em and} an explanation of a recent empirical observation concerning casualty distributions in human insurgencies and terrorism\cite{Bohorquez}. In addition to making microscopic sense for each system in turn, the model successfully reproduces the distributions reported for several other complex phenomena of current interest\cite{Gabaix,Stanley,chialvo,Richardson}.

X-ray diffraction (XRD) studies by Fratini {\it et al.}\cite{Fratini}  in the high-temperature superconducting (SC) cuprate system La$_2$CuO$_{4+y}$ revealed a partial ordering of a stripe-like `$Q2$ superstructure' of the interstitial oxygen atoms (i-Os), depending on the heat treatment of the samples. The probability distribution of the $Q2$ XRD intensity, $x = I(Q2)/I_0$ (normalized to the intensity $I_0$ of the tail of the main crystalline reflections at each spatial point), was shown\cite{Fratini} to exhibit fractal scaling over a wide range: $p(x) \propto x^{-\alpha} {\rm exp}(-x/x_0)$ with $\alpha =2.6 \pm 0.2$ (see Fig. 1(a)). Since the intensity is proportional to the volume of scatterers, this suggests that the distribution of nanoscale volumes of ordered i-Os follows the same power-law pattern. Thus $x$ can be re-interpreted as the volume occupied by $Q2$-ordered i-Os. Remarkably, it was found\cite{Fratini} that this fractal arrangement of $Q2$-ordered i-Os promotes superconductivity in La$_2$CuO$_{4+y}$ (i.e. the SC transition temperature $T_c$ increases when the fractality is more prominent). 

Figure 1(b) shows the prediction of the complexity model which is solved analytically in Appendix A, while the common underlying mechanism is shown schematically in Fig. 2. 
The field of high-$T_c$ superconductivity is packed with a variety of theories\cite{Ashkenazi}, ranging from the conventional BCS approach through to the suggestion of behavior associated with special black holes\cite{Zaanen}. However, as stated in the accompanying description\cite{Zaanen} to Fratini {\it et al.}'s paper\cite{Fratini}, `there is nothing in the textbooks even hinting at an explanation' for why fractal-defect structure, ranging from a micrometer up to fractions of a millimeter scale, should cause the observed enhancement of $T_c$ -- and in particular, why the value\cite{Fratini} $\alpha = 2.6 \pm 0.2$ (Fig. 1(a)) emerges.
 
\begin{figure}  % fig 1
\centering
\includegraphics[width=\textwidth]{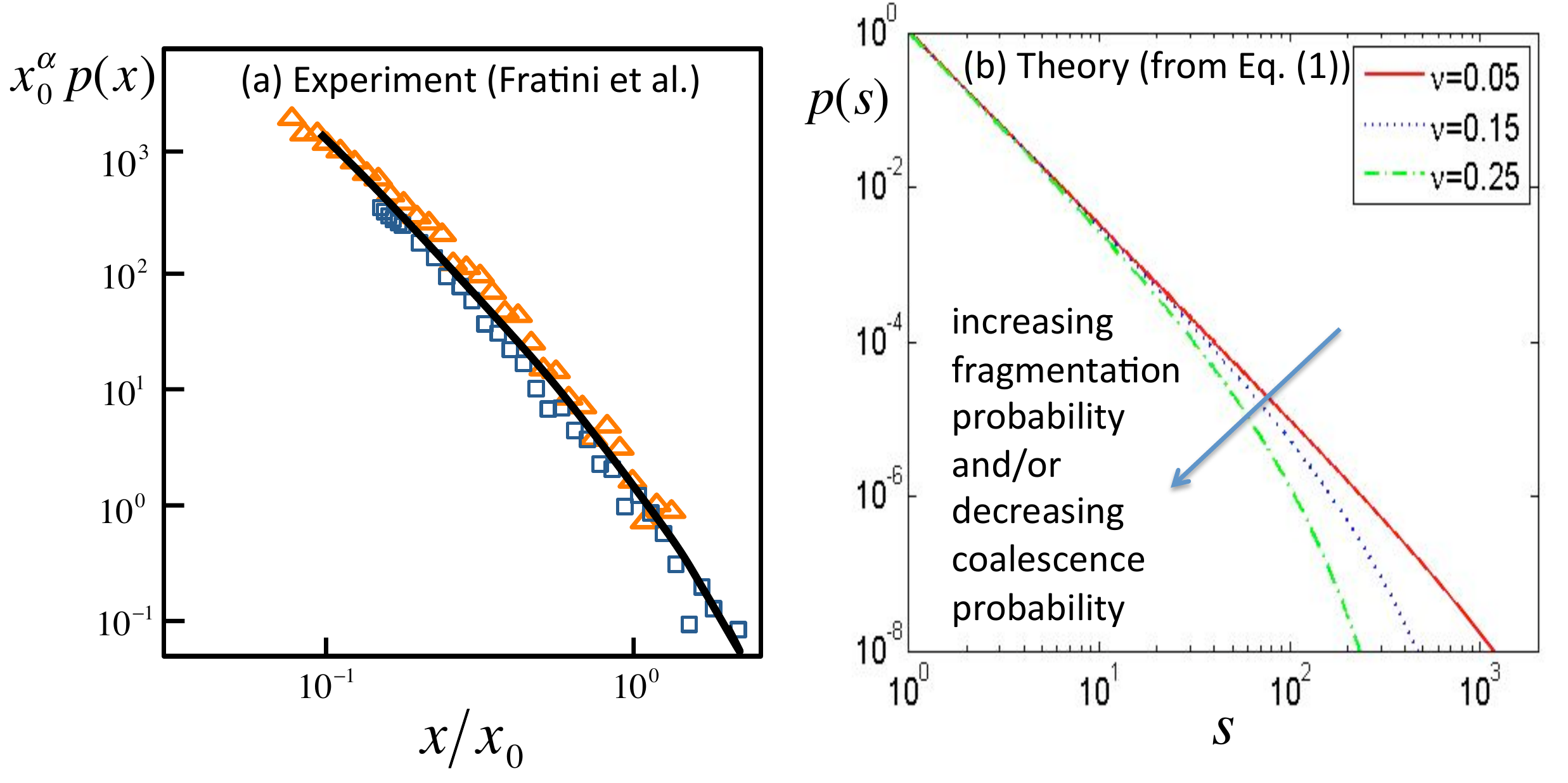}
\caption{(a) Experimental results (triangles and squares) compared to an exponentially truncated power-law curve (black line) having $\alpha =2.6 \pm 0.2$.
Triangles correspond to $T_c=40\;$K and squares correspond to $T_c=16+32\;$K. Information extracted from Fig.~2c of Fratini {\it et al.}\cite{Fratini}, and shows data points lying above the lower cutoff.
(b) Analytic solution of the model (Eq. (1)) corresponds to an exponentially truncated power-law with $\alpha =2.5$ (see Appendix A for full discussion). For illustration purposes, (b) shows $\nu_{\rm frag}=\nu=1-\nu_{\rm coal}$ but the same form emerges for more general values. Following Fratini {\it et al.}\cite{Fratini}, the data in (a) are scaled by $x_0$ and the full exponential tail is not shown. By contrast, (b) shows the full unscaled theoretical form.}
\end{figure}

Experimental evidence\cite{Ong,Alloul,Okada1,Okada2,Dubroka} indicates that cuprate superconductors are characterized by a temperature range $\{T_{\rm fluc}\}$, above $T_c$, where SC fluctuations exist but the global SC phase coherence is incomplete, and that the SC-fluctuation regime does not coincide with the pseudogap phase of the cuprates which probably reflects a two-gap scenario\cite{Deutscher1} (see discussion elsewhere\cite{Ashkenazi}). Furthermore, it has been demonstrated that effects such as compressive strain on thin films\cite{Pavuna} can raise $T_c$ into the SC-fluctuation regime. This regime is therefore characterized by the existence of Cooper pairs (CPs) which lack complete phase coherence throughout the system, with the number of CPs depending on temperature ($T$) in the thermodynamic limit. The ordering of the i-Os in La$_2$CuO$_{4+y}$\cite{Fratini} has a stabilizing effect on the coherence between the CPs which is diminished by the existence of disorder; furthermore, the location of the ordered i-Os between the CuO$_2$ planes is helpful for the establishment of inter-planar coherence. Hence an individual ordered mesoscopic i-O region will provide a favorable environment for the portion of the SC state which finds itself within it, implying that the sample-wide many body state will have spatially heterogeneous coherence depending on how the ordered i-O regions are arranged. Our challenge is therefore to explain how a fractal space with $\alpha = 2.6 \pm 0.2$ manages to protect the many-body phase coherence (and therefore long-range correlations) at temperatures for which an SC state would ordinarily be very fragile.

\begin{figure}  % fig 2
\centering
\includegraphics[width=\textwidth]{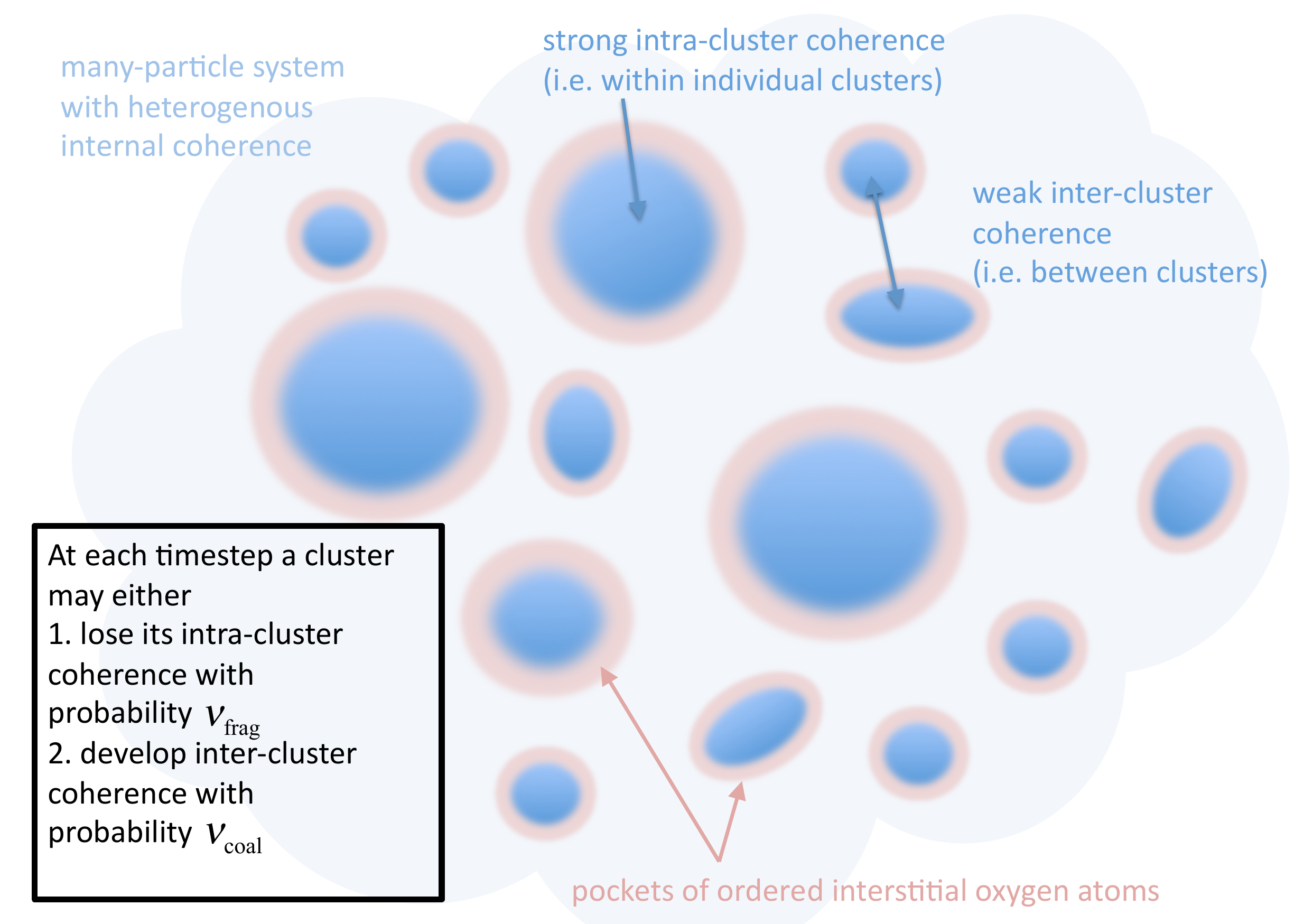}
\caption{Schematic of model processes of coalescence and fragmentation by which coherence of two clusters becomes synchronized (i.e. coalescence) or the coherence within a single cluster is suddenly destroyed by a decohering event (i.e. fragmentation). These processes are generic, but the language chosen is for specific application to superconductivity.}
\end{figure}

Irrespective of the actual microscopic pairing mechanism giving rise to coherent CPs, one can therefore model the $\{T_{\rm fluc}\}$ regime as a dynamical competition between (1) the entropy-driven tendency of a coherent SC state to break up (i.e. fragment) through phase-breaking events, into smaller clusters where the intra-cluster phase coherence is strong but the inter-cluster coherence is weak; and (2) the tendency of the underlying SC mechanism to synchronize the phases within two clusters, forming one larger coherent droplet as shown schematically in Fig. 2. Close to $T_c$, (2) will dominate, while considerably above it (1) will dominate -- however both tendencies coexist for $T \in \{T_{\rm fluc}\}$. 
Hence a $T$-dependent population of $N \gg 1$ CPs comprises, for $T \in \{T_{\rm fluc}\}$, a set of clusters with each one having its own internal phase coherence. Each cluster has a size $s$ representing the number of phase-coherent objects (i.e. CPs), and the number $n_s(t)$ of such clusters evolves over time as a result of fragmentation and coalescence (i.e. (1) and (2) above) such that $\sum_{s=1}^{N} s n_s(t) = N$ at each timestep. At each timestep, a coherence-breaking event will occur with probability $\nu_{\rm frag}$ in a cluster
which is randomly chosen according to its size $s$, mimicking the
fact that larger clusters will suffer proportionally more quantum fluctuations. In short, a cluster with more members has more chances of initiating a dephasing event. With probability $\nu_{\rm coal}$, this cluster instead coalesces with another one chosen randomly according to its size. This mimics the tendency of two clusters to synchronize their phase coherence as the system heads toward global phase-coherence in the ideal SC state, i.e. any subsequent coalescence events will likely be initiated by pairwise coherence between individual members in the two groups and hence the probability will depend on the number of members.  We stress that we use the term `coalescence' to simply mean that two groups act in a coordinated way, not necessarily that they are physically joined. As $T$ is lowered towards $T_c$, $\nu_{\rm frag}$ is likely to decrease due to the reduced frequency of coherence-breaking processes, while $\nu_{\rm coal}$ is likely to increase as the establishment of a coherent SC state is approached. 

The Master Equations for this model are:
\begin{eqnarray}
\frac{\partial n_s}{\partial
t}&=&{\frac{\nu_{\rm coal}}{N^2}\sum^{s-1}_{k=1}kn_k(s-k)n_{s-k}}-{\frac{\nu_{\rm frag}sn_s}{N}}
-{\frac{2\nu_{\rm coal}sn_s}{N^2}\sum^{\infty}_{k=1}kn_k}
\ , \quad s\geq2  \ , \label{eq:genez0}\\
 \frac{\partial n_1}{\partial t}&=&\frac{\nu_{\rm frag}}{N}\sum^{\infty}_{k=2}k^2n_k-\frac{2\nu_{\rm coal}n_1}{N^2}\sum^{\infty}_{k=1}kn_k \ , \nonumber
\end{eqnarray}
where we make the sensible approximation that $N\rightarrow\infty$. Simpler versions of Eq. (1) have been presented before\cite{hulst,Johnson,Eguiluz,Ruszczycki}, however this is the first treatment with general values for $\nu_{\rm frag}$ and $\nu_{\rm coal}$ and it is the first application to superconductivity. The distance-independence is justified by the fact that the phase coherence in a macroscopic SC state can exist over large distances (e.g. $10^{-9}$m-$10^{-4}$m) even if it is fragile (Fig. 2), and hence makes Eq. (1) equivalent to a mean-field model (i.e. no distance dependence). Solving the model analytically in the steady-state limit (i.e. 
${\partial n_s}/{\partial t}=0$) yields a closed-form expression for the time-averaged probability distribution for phase-coherent cluster sizes which is in exact agreement with Fratini {\it et al.}\cite{Fratini}, i.e. it yields an exponentially truncated power-law $p(s) \propto s^{-\alpha} {\rm exp}(-s/s_0)$ with $\alpha = 5/2\equiv 2.5$ (see Appendix A). We also uncover the following novel theoretical relationship between cutoff $s_0$ and fragmentation and coalescence probabilities $\nu_{\rm frag}$ and $\nu_{\rm coal}$:
\begin{equation}
s_0=-\left[{\rm ln} \left(\frac{4(\nu_{\rm frag}+\nu_{\rm coal})\nu_{\rm coal}}{(\nu_{\rm frag}+2\nu_{\rm coal})^2}\right)\right]^{-1}\ \ .
\end{equation}

\begin{figure} % fig 3 
\centering
\includegraphics[width=\textwidth]{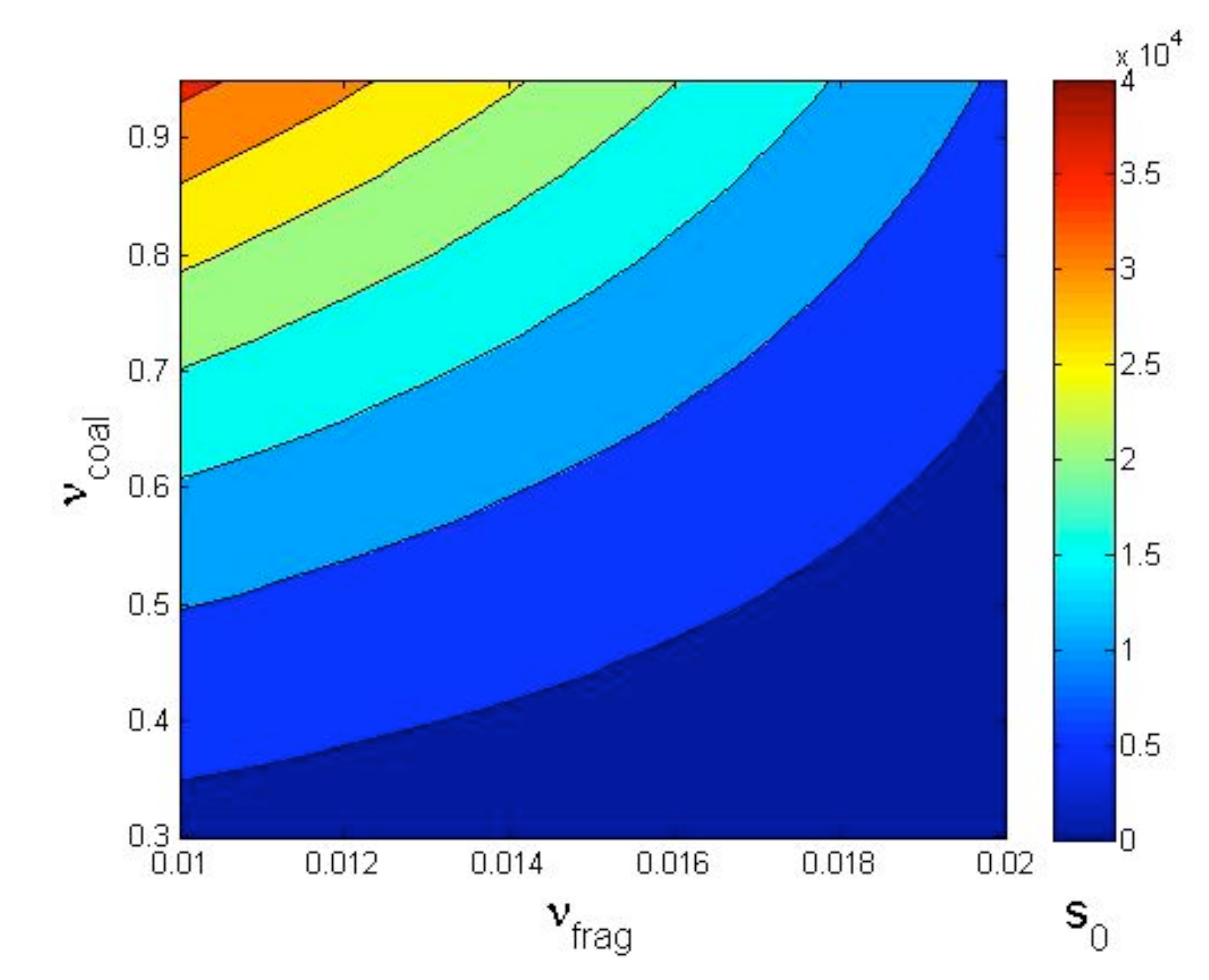}
\caption{Model prediction (Eq. (2)) for how the exponential cutoff $s_0$, the  fragmentation probability $\nu_{\rm frag}$ and the coalescence probability $\nu_{\rm coal}$ are related. A portion of the possible parameter value space is shown for illustration.}
\end{figure}

The model's $5/2$ truncated power-law form is remarkably robust to generalizations, e.g. coalescence of multiple groups, fragmentation into groups larger than one, a slowly time-varying particle number $N$, and it holds for a wide range of $\nu_{\rm frag}$ and $\nu_{\rm coal}$ values\cite{Ruszczycki,Eguiluz,hulst,Johnson}. It does not depend on the size of the individual fragmented parts, as long as they are all small, since the size of the largest pieces just dictates the value of $s$ above which the $5/2$ power-law result kicks in. Figure 1(b) illustrates the model's predicted analytic form, for the simple demonstrative case of $\nu_{\rm frag}=\nu=1-\nu_{\rm coal}$. As $\nu_{\rm frag}$ increases and/or $\nu_{\rm coal}$ decreases, the cutoff $s_0$ sets in at lower $s$ and eventually dominates the power-law. (For Fig. 1(a), adapted from Fratini {\it et al.}\cite{Fratini}, the axis rescaling shifts the red experimental curve to the right and the tails fall outside the range shown, as opposed to Fig. 1(b) which shows the full unscaled theoretical form including the exponential tail). Equation (2) is illustrated in Fig. 3. In addition to reproducing Fratini {\it et al.}'s\cite{Fratini} precise functional form, the model allows us to interpret the cutoff $s_0$ in terms of the dynamic processes of fragmentation and coalescence (see Fig. 3). 

Our theory does not require any specific behavior at the level of individual CPs, nor precise details or mechanisms involving the location of the i-Os between the CuO$_2$ planes. Instead, it takes the novel approach of simultaneously describing the partially coherent SC state across several orders of lengthscale magnitude, predicting that it can fit perfectly into the fractal space created by the ordered i-Os such that each cluster resides within a pocket full of ordered i-Os, as sketched in Fig. 2. Since the fractal space of the ordered i-Os (which develops in the samples at high temperatures where the heat treatment is applied\cite{Fratini}) is basically static for $T \in \{T_{\rm fluc}\}$, the coupling of the fractal space of the CPs to it results in an increase of SC coherence (see above). Consequently $T_c$ is raised within the $\{T_{\rm fluc}\}$ temperature regime, as has been observed\cite{Fratini}. In samples where the fractal i-O box is very incomplete, $T_c$ is not raised throughout an entire sample which hence results in the observed mixed state\cite{Fratini}. 

A question to be clarified in future research is whether the ordered i-O cluster formation process follows a similar mechanism to that of the CP cluster formation. Similarly to the CPs, the collective (phonon-like) excitations, associated with the ordering of the i-Os, have a quantum nature and can in principle exhibit long-range phase coherence. The mobility of the i-Os is quite high at the heat-treatment temperatures\cite{Fratini}, and fluctuations of the $Q2$-ordering phase transition occur. However, the ordering phase transition cannot be completed throughout the sample since the mobility is reduced when $T$ is lowered -- hence an i-O fractal structure gets locked in at low $T$.  

The Master Equations in Eq. (1) also provide a plausible explanation for the recently observed tendency of modern insurgent groups to produce approximate power-law casualty distributions with\cite{Bohorquez} $\alpha\approx 2.5$. In this human context, the distance-independent coalescence of Eq. (1) mimics the availability of modern electronic communication between clusters and hence the possibility of some long-range inter-cluster `coherence', while the fragmentation mimics the tendency for an individual cluster to scatter when suddenly sensing danger and hence lose its intra-cluster coherence\cite{Bohorquez} (see Appendix A). The process of superconductivity enhancement and human conflict are therefore dynamically equivalent in that they can both be approximated by Eq. (1), and hence both produce identical exponentially truncated 2.5 power-law forms. We stress that we are not saying that these systems are {\em physically} identical, since obviously both the objects involved and the origins of their correlations are very different  -- but they do behave dynamically {\em as if} they were both the same system (i.e. they both follow Eq. (1)).

An immediate practical consequence is that learning how to `protect' a population under attack from an insurgent or terrorist threat, and learning how to `protect' a fragile SC state from decoherence and thus raise $T_c$, become inter-related problems at this Master Equation level. Hence insights from one might help the other. This moves the Complexity debate forward from the issue of whether systems are physically similar or not, to whether they are dynamically equivalent or not at some given level of approximation (e.g. Master Equations for cluster sizes). To illustrate that it is not just SC systems that can belong to a given complexity universality class, we note that fragmentation in low-dimensional Bose disordered systems was recently identified as a novel mechanism for detecting a superfluid-insulator quantum phase transition\cite{savona}.
 
This model (Eq. (1)) also offers an alternative explanation for a variety of other complex phenomena which have been found to exhibit a robust 2.5 power-law. Gabaix {\it et al.}\cite{Gabaix} found a common power-law distribution for individual transaction sizes with $\alpha=2.5\pm 0.1$, for the London Stock Exchange, the NYSE, and the Paris Bourse. Interpreting $N$ as the average aggregate demand for stocks, this demand $N$ gets shaped into a distribution of demand `clusters' representing potential orders of a given size $s$. 
Since it is reasonable to expect orders to be realized at random, the distribution of individual transaction sizes is proportional to the distribution of clusters of potential orders -- hence $\alpha = 2.5$. Similarly, Richardson\cite{Richardson} concluded that the distribution of approximately $10^3$ gangs in Chicago, and in Manchoukuo in 1935, separately followed a truncated power-law with $\alpha \approx 2.3$. Interpreting $N$ as the number of potential gang members in each case, with each comprising a transient soup of clusters which tend to combine or fragment over time, yields $\alpha = 2.5$. In a similar way, the robust time-dependence of a power-law with $\alpha\approx 2.4$ in a recent New York garment industry study\cite{felix} can be reinterpreted as a repartitioning of trading interactions, with multi-component clusters continually being built up as part of common jobs (i.e. coalescence) and then dissolving upon completion (i.e. fragmentation). 
For collections of $N$ neurons\cite{chialvo}, we can imagine a dynamical coalescence-fragmentation grouping process in which groups of neurons become synchronized, and then this synchronization ultimately fragments. (Members of the same group need not be physically adjacent to each other). When an entire group fires, it creates a measurable activity equal to\cite{chialvo} the group size $s$. Hence the resulting activity distribution will follow a new power-law given by $s\times p(s)$. The resulting power-law exponent $(\alpha-1) = 1.5$, which is exactly the famous empirical $3/2$ value\cite{chialvo}. 
We note that although competing theories exist for many of these applications (i.e. Chialvo\cite{chialvo} for neural dynamics and Gabaix {\it et al.}\cite{Stanley} for markets), we know of no other single mechanism which is simultaneously physically plausible for each application area {\em and} which can explain the mysterious finding of Fratini {\it et al.}\cite{Fratini}. 

Additionally, the model yields several concrete predictions for superconductivity in the cuprates, and similar quantum phenomena, which we hope will stimulate future empirical investigation. First, it should be possible to manipulate host material properties such as to vary $\nu_{\rm frag}$ and/or $\nu_{\rm coal}$ and hence alter the cutoff value $s_0$ in Eq. (2). In particular, if $s_0$ is measured across a wide range of samples, it should be possible to infer best-fit values for the {\em dynamical} quantities $\nu_{\rm frag}$ and $\nu_{\rm coal}$ which directly affect the SC state's coherence, and which might have been impossible to estimate by other means. 
Second, in the limit where the spatial extent of any potential phase coherence (i.e. cluster coalescence) is small, the model effectively becomes a low dimensional percolation problem in which the truncated power-law form now appears only at the percolation threshold $p_c$\cite{Johnson}. As we move from the infinite dimensional (i.e. long-range interaction) model in Eq. (1) toward a two-dimensional limit with only nearest neighbor interactions, $\alpha \rightarrow 2$ at $p_c$\cite{Johnson}. This behavior has been analyzed by Laibowitz {\it et al.}\cite{Deutscher2} using a similar two-dimensional model, to account for the observed cluster statistics in thin films near $p_c$.
Away from $p_c$, the distribution then completely loses its power-law characteristics, which is exactly what we have observed in the transition from modern insurgent wars (where long-range interactions are possible through communication devices and the Internet) to older wars which resemble fights on a grid and in which interactions and information transfer are effectively limited to nearest neighbors. It would be fascinating to explore this same transition in La$_2$CuO$_{4+y}$ (or an analogous SC system) by manipulating the fractal distribution of ordered i-Os during the annealing process.
Third, it has been recently established that the model's {\it temporal} evolution is able to reproduce the large fluctuations associated with\cite{Zhao} human contagion in financial markets (currency trading on the one minute timescale), biological systems (cold transmission in schools), and social systems (YouTube downloads). It would be interesting to see if the SC fluctuations near $T_c$ are also well described by the stochastic numerical simulations of the model's temporal dynamics, and whether the averaged behavior of individual groups in the SC state near $T_c$ ends up following the analytic Master equations. Applying ultrafast optical pump-probe measurements on the SC state, it should be possible to track the fragmentation process and associated bursty behavior as a large phase-coherent cluster fragments and smaller clusters take over. 
Fourth, our finding raises the possibility that other types of quantum phase transition might also benefit from a space which better matches the possible fractal form of the underlying quantum many-body state. In particular, the survival of other exotic, yet fragile, quantum many-body states might be similarly enhanced by incorporating a protective 2.5 power-law `box' within the host material microstructure. Such states may include highly nontrivial quantum correlations beyond two-body entanglement\cite{vedral}. Indeed, a measurable shift in transition temperatures could be a novel, noninvasive way of probing such multipartite entanglement\cite{vedral,luis,Vedral2,Simmons}. 

%\newpage

%{\bf \noindent Appendix}

\appendix

\section{}

Here we present the derivation of the exponentially cutoff 2.5 power-law. Analysis of a simpler version of Eq. (1), was completed by d'Hulst and Rodgers\cite{hulst}, and real-world applications have focused on financial markets -- however this is the first example of an application to superconductivity and derivation with general values $\nu_{\rm frag}$ and $\nu_{\rm coal}$. At each timestep, the internal coherence of a population of $N$ objects (which we refer to as an `agents' to acknowledge possible application to human systems) comprises a heterogenous soup of clusters. Within each cluster, the component objects have a strong intra-cluster coherence. Between clusters, the inter-cluster coherence is weak. An agent $i$ is then picked at random -- or equivalently, a cluster is randomly selected with probability proportional to size. Let $s_i$ be the size of the
cluster to which this agent belongs. With probability $\nu_{\rm frag}$, the coherence of a given cluster fragments completely into $s_i$ clusters of size one. If it doesn't fragment, a second cluster is randomly selected with probability again proportional to size -- or equivalently, another agent $j$ is picked at random. With probability $\nu_{\rm coal}$, the two groups then coalesce (or develop a common `coherence' in the case of the superconducting application). 
The Master Equation is as
follows:
\begin{eqnarray}
\frac{\partial n_s}{\partial
t}&=&{\frac{\nu_{\rm coal}}{N^2}\sum^{s-1}_{k=1}kn_k(s-k)n_{s-k}}-{\frac{\nu_{\rm frag}sn_s}{N}}-{\frac{2\nu_{\rm coal}sn_s}{N^2}\sum^{\infty}_{k=1}kn_k}
\ , \quad s\geq2  \ , \label{eq:genez1}\\
 \frac{\partial n_1}{\partial t}&=&\frac{\nu_{\rm frag}}{N}\sum^{\infty}_{k=2}k^2n_k-\frac{2\nu_{\rm coal}n_1}{N^2}\sum^{\infty}_{k=1}kn_k \ . \label{eq:genez2}
\end{eqnarray}
Note here we make an approximation that $N\rightarrow\infty$. The
terms on the right hand side of Eq.~(\ref{eq:genez1}) represent
all the ways in which $n_s$ can change.
In the equilibrium state:
\begin{eqnarray}
sn_s&=&\frac{1-\nu_{\rm frag}}{(\nu_{\rm frag}+2\nu_{\rm coal})N}\sum^{s-1}_{k=1}kn_k(s-k)n_{s-k}
\ ,
\quad s\geq2 \ , \label{eq:ezeq}\\
n_1&=&\frac{\nu_{\rm frag}}{2(1-\nu_{\rm frag})}\sum^{\infty}_{k=2}k^2n_k\
\ . \label{eq:ezeq1}
\end{eqnarray}
Consider
\begin{equation}
G[y]=\sum^{\infty}_{k=0}kn_ky^k=n_1y+\sum^{\infty}_{k=2}kn_ky^k
\equiv n_1y+g[y] \ , \label{eq:generating}
\end{equation}
where $y$ is a parameter and g[y] governs the cluster size
distribution $n_k$ for $k \geq 2$. Multiplying Eq.~(\ref{eq:ezeq})
by $y^s$ and then summing over $s$ from $2$ to $\infty$, yields:
\begin{equation}
g[y]=\frac{1-\nu_{\rm frag}}{(\nu_{\rm frag}+2\nu_{\rm coal})N}G[y] \ ,
\end{equation}
i.e.
\begin{equation}
g[y]^2-\left(\frac{\nu_{\rm frag}-2\nu_{\rm coal}}{\nu_{\rm coal}}N-2n_1y\right)g[y]+n_1^2y^2=0\label{eq:g[y]}
\ .
\end{equation}
From Eq.~(\ref{eq:generating}), $g[1]=G[1]-n_1$. Substituting this
into Eq.~(\ref{eq:g[y]}) and setting $y=1$, we can solve for
$g[1]$
\begin{equation}
g[1]=\frac{\nu_{\rm coal}}{\nu_{\rm frag}+2\nu_{\rm coal}}N \ .
\end{equation}
Hence
\begin{equation}
n_1=N-g[1]=\frac{\nu_{\rm frag}+\nu_{\rm coal}}{\nu_{\rm frag}+2\nu_{\rm coal}}N \
.
\end{equation}
Substituting this into Eq.~(\ref{eq:g[y]}) yields
\begin{equation}
g[y]^2-\left(\frac{\nu_{\rm frag}+2\nu_{\rm coal}}{\nu_{\rm coal}}N-\frac{2N(\nu_{\rm frag}+\nu_{\rm coal})}{\nu_{\rm frag}+2\nu_{\rm coal}}y\right)g[y]+\frac{(N(\nu_{\rm frag}+\nu_{\rm coal}))^2}{(\nu_{\rm frag}+2\nu_{\rm coal})^2}y^2=0
\ .
\end{equation}
We can solve this quadratic for $g[y]$
\begin{equation}
g[y]=\frac{(\nu_{\rm frag}+2\nu_{\rm coal})N}{4\nu_{\rm coal}}\left(2-\frac{4(\nu_{\rm frag}+\nu_{\rm coal})\nu_{\rm coal}}{(\nu_{\rm frag}+2\nu_{\rm coal})^2}y-2\sqrt{1-\frac{4(\nu_{\rm frag}+\nu_{\rm coal})\nu_{\rm coal}}{(\nu_{\rm frag}+2\nu_{\rm frag})^2}y}\right)
\ ,
\end{equation}
which can be easily expanded
\begin{equation}
g[y]=\frac{(\nu_{\rm frag}+2\nu_{\rm coal})N}{2\nu_{\rm coal}}\sum^{\infty}_{k=2}\frac{(2k-3)!!}{(2k)!!}\left(\frac{4(\nu_{\rm frag}+\nu_{\rm coal})\nu_{\rm coal}}{(\nu_{\rm frag}+2\nu_{\rm coal})^2}y\right)^k
\ .
\end{equation}
Comparing with the definition of $g[y]$ in
Eq.~(\ref{eq:generating}) shows that
\begin{equation}
n_s=\frac{\nu_{\rm frag}+2\nu_{\rm coal}}{2\nu_{\rm coal}}\frac{(2s-3)!!}{s(2s)!!}\left(\frac{4(\nu_{\rm frag}+\nu_{\rm coal})\nu_{\rm coal}}{(\nu_{\rm frag}+2\nu_{\rm coal})^2}\right)^s
\ .
\end{equation}
We now employ Stirling's series
\begin{equation}
ln[s!]=\frac{1}{2}ln[2\pi]+\left(s+\frac{1}{2}\right)ln[s]-s+\frac{1}{12s}-...
\ .
\end{equation}
Hence for $s\geq2$, we find
\begin{equation}
n_s\approx\left(\frac{(\nu_{\rm frag}+2\nu_{\rm coal})e^2}{2^{3/2}\sqrt{2\pi}\nu_{\rm coal}}\right)\left(\frac{4(\nu_{\rm frag}+\nu_{\rm coal})\nu_{\rm coal}}{(\nu_{\rm frag}+2\nu_{\rm coal})^2}\right)^s\frac{(s-1)^{2s-3/2}}{s^{2s+1}}N
\ ,
\end{equation}
which implies that
\begin{equation}
n_s \sim
\left(\frac{\nu_{\rm coal}^{s-1}(\nu_{\rm frag}+\nu_{\rm coal})^s}{(\nu_{\rm frag}+2\nu_{\rm coal})^{2s-1}}\right)s^{-5/2}\ \ . \label{eq:power}
\end{equation}
In the limit $s\gg 1$, this is formally equivalent to saying that
\begin{equation}
n_s \sim
{\rm exp}(-s/s_0) s^{-5/2}\label{eq:power2}
\end{equation}
where
\begin{equation}
s_0=-\left[{\rm ln} \left(\frac{4(\nu_{\rm frag}+\nu_{\rm coal})\nu_{\rm coal}}{(\nu_{\rm frag}+2\nu_{\rm coal})^2}\right)\right]^{-1}\ \ .
\end{equation}
For large cluster sizes (i.e. large $s$ such that $s\sim O(N)$) the power law behaviour is masked by the
exponential function. The equilibrium state for the distribution
of cluster sizes can therefore be considered a power-law with
exponent $\alpha\sim2.5$, together with an exponential cut-off.
The exponent $\alpha$ can be manipulated by implementing
suitably chosen microscopic rules, as discussed by us and others elsewhere.

In the human context, the fact that the interactions are effectively distance-independent as far as Eq. (1) is concerned, captures the fact that we wish to model systems where messages can be transmitted over arbitrary distances (e.g. modern human communications). Bird calls and chimpanzee interactions in complex tree canopy structures can also mimic this setup, as may the increasingly longer-range awareness that arises in larger animal, fish, bird and insect groups. In a human/biological context, a justification for choosing a cluster with a probability which is proportional to its size, is as follows: a cluster with more members has more chances of initiating an event. In the materials context, it will have a higher chance of feeling an interaction from a dephasing mechanism event (e.g impurity or phonon) and hence may fragment more readily. It will also be more likely to find members of another cluster more frequently, and hence be able to synchronize with them -- thereby synchronizing the two clusters. 
It is well documented that clusters of living objects (e.g. animals, people) may suddenly scatter in all directions (i.e. complete fragmentation as in Eq. (1)) when its members sense danger, simply out of fear or in order to confuse a predator. Such fleeing behavior was discussed at length in the classic 1970 work `Protean Defence by Prey Animals' by D. A. Humphries and P.M. Driver, Oecologia (Berl.) {\bf 5}, 285-302 (1970). Clusters of inanimate objects such as doubly-ionized Argon atoms and animal Hox genes, also exhibit complete fragmentation.

\end{document}